\documentclass[manuscript]{aastex61}
\usepackage{natbib}
\usepackage{graphicx}

\newcommand{\hipp}{{\it Hipparcos}\,\,}
\newcommand{\masyr}{\,mas\,${\rm yr}^{-1}$\,}

\received{receipt date}
\revised{revision date}
\accepted{acceptance date}
\published{published date}
\submitjournal{AJ}

\shorttitle{Betelgeuse's Parallax}
\shortauthors{Harper et al.}

\begin{document}


\title{An Updated 2017 Astrometric Solution for Betelgeuse}


\author{G. M. Harper}
\author{A. Brown}
\affiliation{Center for Astrophysics and Space Astronomy, University of Colorado,
    Boulder, CO 80309, USA}
\email{graham.harper@colorado.edu}

\author{E. F. Guinan}
\affiliation{Villanova University, PA 19085, USA}

\author{E. O'Gorman}
\affiliation{Dublin Institute for Advanced Studies, Dublin 2, Ireland}

\author{A. M. S. Richards}
\affiliation{Jodrell Bank Centre for Astrophysics, School of Physics and Astronomy, University of Manchester, UK}

\author{P. Kervella}
\affiliation{Unidad Mixta Internacional Franco-Chilena de Astronom\'{i}a (UMI 3386), Departamento de Astronom\'{i}a, Universidad de Chile, Santiago, Chile.} 
\affiliation{Observatoire de Paris, PSL Research University, CNRS, UPMC, Univ. Paris-Diderot, France.}

\author{L. Decin}
\affiliation{Instituut voor Sterrenkunde, Katholieke Universiteit Leuven, Celestijnenlaan 200D, 3001 Leuven, Belgium}



\begin{abstract}

We provide an update for the astrometric solution for the Type II supernova progenitor Betelgeuse using the
revised {\it Hipparcos} Intermediate Astrometric Data (HIAD) of van Leeuwen, combined with existing VLA and new e-MERLIN and ALMA positions. The 2007 {\it Hipparcos} refined abscissa measurements required the addition of so-called {\it Cosmic Noise} of 2.4\,mas to find an acceptable 5-parameter stochastic solution. We find that a measure of radio {\it Cosmic Noise} should also be included for the radio positions because surface inhomogeneities exist at a level significant enough to introduce additional intensity centroid uncertainty. Combining the 2007 HIAD with the proper motions based solely on the radio positions leads to a parallax of $\pi = 5.27\pm 0.78$\,mas ($190^{+33}_{-25}$\,pc), smaller than the \hipp 2007 value of $6.56\pm 0.83$\,mas ($152^{+22}_{-17}$\,pc; van Leeuwen 2007). Furthermore, combining the VLA and new e-MERLIN and ALMA radio positions with the 2007 HIAD, and including radio {\it Cosmic Noise} of 2.4\,mas, leads to a nominal parallax solution of $4.51 \pm 0.80$\,mas 
($222^{+48}_{-34}$\,pc), which while only $0.7\sigma$ different from the 2008 solution of Harper et al. it is $2.6\sigma$ different from the solution of van Leeuwen.
An accurate and precise parallax for Betelgeuse is always going to be difficult to obtain because it is small compared to the stellar angular diameter ($\theta=44$\,mas). We outline an observing strategy, utilizing future mm and sub-mm 
high-spatial resolution interferometry that must be used if substantial improvements in precision and accuracy of the parallax and distance are to be achieved.

\end{abstract}


\keywords{stars: astrometry --- stars: individual($\alpha$~Ori)}



\section{Introduction}

The parallax of the red supergiant Betelgeuse ($\alpha$~Ori: M2~Iab) is of considerable interest to the astrophysics community because it is required to determine the fundamental stellar parameters, and to constrain its evolutionary status, e.g., an age of $\sim 8-10$\,Myr, with $<1$\,Myr until it explodes \citep{2008AJ....135.1430H, 2016ApJ...819....7D}. There is a paucity of M supergiants in the solar neighborhood, and Betelgeuse is one of only two nearby supernova progenitors - the other being Antares (M1~Iab + B3~V: dist$\simeq 200$\,pc).
If Betelgeuse explodes as a red supergiant Betelgeuse will likely become a Type II-P, or if it evolves blueward it may become a Type II-L, see \citet{2009AJ....137.3558S}.
As a bright large angular diameter source it has been the focus of innumerable multi-wavelength observational studies, but relatively few theoretical ones. It is frustrating that its parallax is currently so poorly constrained: $5.07\pm 1.10(1\sigma)$\,mas ($197^{+55}_{-35}$\,pc; Harper et al. 2008), which severely limits what can be gleaned from all the research effort into this system. The luminosity reflects the star's mass and, hence, its evolution and lifetime, and with its interferometeric angular diameter of $\simeq 44$\,mas, 
\cite[e.g.,][]{O2009A&A...503..183O, H2009A&A...508..923H, 2016A&A...588A.130M}, a direct estimate of the surface
gravity can be obtained. Accurate stellar parameters are also crucial to limit the parameter space of theoretical and numerical studies of stellar structure, \citep[e.g.,][]{2002AN....323..213F}, mass loss mechanism \citep{1984ApJ...284..238H, A2000ApJ...528..965A}, and circumstellar structure \citep{2017ApJ...836...22H}.

The inherent problem for astrometric solutions for stars with angular diameters larger than their parallax is that any deviation of the center of intensity from the
center-of-mass (CoM or barycenter) leads to additional systematic uncertainties. Both Betelgeuse and Antares ($\alpha$~Sco:M1~Iab) have angular diameters, $\simeq 44$~mas, and distances near 200~pc and their parallaxes are $\simeq 5$\,mas. The release of the original \hipp catalog \citep{1997A&A...323L..49P,ESA1997} suggested that Betelgeuse had a distance of $131^{+35}_{-23}$\,pc, closer than many other astrophysical estimates, \citep[see][and references therein]{2008AJ....135.1430H}.
That astrometric solution required the addition of $3.4\pm 0.6$\, mas of {\em Cosmic Noise}\footnote{A stochastic solution is one in which the dispersion of the abscissa residuals about the optimum solution was greater than expected and an additional source of {\em Cosmic Noise (or Dispersion)} was added to the residual uncertainties.} to obtain a stochastic 5-parameter astrometric solution
[i.e., R.A. ($\alpha$), decl. ($\delta$), parallax ($\pi$), and proper motions in $\alpha\cos\delta$ and $\delta$]. The specific origin of this additional positional uncertainty remains unknown: potential origins include source size, brightness, instrumental effects, and movement of the stellar
photo-center which may result from large-scale convective/pulsation-induced motions in the outer layer of the star
\citep[e.g.,][]{2002AN....323..213F}. \citet{B2003AJ....126..484B} derived somewhat different proper-motions from the 1997 \hipp solution, and this was confirmed by subsequent multi-wavelength VLA cm-radio interferometry \citep{2008AJ....135.1430H}. This latter study combined VLA radio positions with the 1997 HIAD and found
a greater distance ($197^{+55}_{-35}$\,pc) and it also revealed a tension whereby the radio positions alone suggested a greater distance. 

Just at the completion of the \citet{2008AJ....135.1430H} study, revised \hipp astrometry was published by 
\citet[hereafter referred to as the 2007 \hipp solution]{2007ASSL..350.....V}.  The astrometric solutions from this impressive re-analysis are generally of significantly improved quality compared to the 1997 release, and the number of single star astrometric solutions requiring {\em Cosmic Noise} was reduced from 1561 to 588. Betelgeuse's revised \hipp parallax put it at $152\pm 20$\,pc, but the solution still required 2.4\,mas of {\em Cosmic Noise}.  
Note that Antares also required 3.6\,mas of {\em Cosmic Noise}.
Recalling that the parallax derived by \citet{2008AJ....135.1430H} was based on the combination of
VLA radio positions and the \hipp Intermediate Astrometric Data 
\citep[HIAD]{1997A&A...323L..49P, ESA1997}, we have been asked whether the revised \citet{2007ASSL..350.....V} intermediate astrometric data would lead to a different
VLA+\hipp combined solution. Here we address this question and provide updated astrometric solutions.

The importance of the radio positions for the astrometric solution is the long timeline that they provide when compared to the \hipp mission. While \hipp observed the sky for 3.2 years, Betelgeuse was observed for less than 28\, months. In contrast,  the radio time-line now extends from 1982 to 2016 ($\Delta t = 34$\,yr), and since 2008 Betelgeuse has also been observed with e-MERLIN \citep{2016A&G....57c3.28G} by \citet{2013MNRAS.432L..61R} and ALMA. The yearly proper motions of Betelgeuse  are $\simeq 10-25$\masyr, which is a significant fraction of the angular diameter per year, so the long timeline helps to define the proper-motion elements of the astrometric solution even with modest measurement uncertainties.

In \S2 we present the new radio positions, discuss the magnitude of expected {\em Cosmic Noise} for the radio positions, and derive radio-only proper motions. In \S3 we give the combined radio and 2007 \hipp astrometric solutions, and in \S4 we discuss what is needed to find a significantly improved parallax for Betelgeuse. Conclusions are given in \S5.

\section{New Radio Positions}

The new radio positions, which consist of two ALMA observations at 330-345\,GHz and three e-MERLIN observations obtained at 5.75\, GHz (5.2\,cm) \citep{2013MNRAS.432L..61R}, are given in Table~\ref{tab:positions}.

\subsection{ALMA}

Betelgeuse was observed by ALMA in a long baseline configuration on 2015 November 9 and again on 2016 August 16 in a more compact configuration (Project code: 2015.1.00206.S , PI: P. Kervella). Both sets of observations were carried out in Band 7 with identical spectral setups and will be described in more detail in a future publication (P. Kervella, et al. 2017, in preparation) For the purpose of this work, the channels containing line emission were excluded from the analysis and a single continuum data set centered at $\simeq$338 GHz with a $\sim$5.9\,GHz bandwidth was used. The maximum baseline in the long baseline observation was 16 km, which yielded an angular resolution of $\sim$15\,mas, while the maximum baseline provided by the more compact configuration observation was 1.2 km and yielded an angular resolution of 185\,mas. The observations of Betelgeuse (Star) were interleaved with observations of the compact phase calibrator J0552+0313 (PhaseCal) located approximately 4\,deg away from the target using a 2-min cycle time. The compact source J0605+0939 (ChkSource) located approximately 7 deg from the PhaseCal was used as a check source. ChkSource was found to be located 3.3\,mas away from its expected position. For the compact configuration the same phase calibrator (J0552+0313) was again used but a longer cycle time of 7.5 min was implemented. J0603+0622 was used as a check source and was found to be located 15 mas away from its expected position. For both epochs, the source position was extracted from the un-self-calibrated continuum datasets using the Python-based task uvmultifit 
\citep{MV2014A&A...563A.136M} and fitting a uniform intensity elliptical disk to the calibrated visibilities.

The positional uncertainties ($Unc$) for the two Betelgeuse observations are dominated by the phase-referencing uncertainty (and not fit errors). We used the difference between the measured positions of ChkSource from those expected, 
$\Delta{\rm CS}$, scaled by the relative separation of Star and PhaseCal, and the 
ChkSource and PhaseCal, i.e.,  
$$Unc = \Delta{\rm CS}\times({\rm Star-PhaseCal})/({\rm ChkSource-PhaseCal})$$
These uncertainties for the ALMA positions are given in Table~\ref{tab:propermotion_alt}. During the compact configuration
observation, the wind speed was high and increasing during the observations, which increased the atmospheric instability. This in turn might lead to an underestimation of the uncertainty reported in Table~\ref{tab:propermotion_alt}.

\subsection{e-MERLIN}

Betelgeuse was observed in the first semester of e-MERLIN  open time in 2012 July, with a bandwidth of 512~MHz
centered on 5.75\,GHz (5.2\,cm). Seven antennas were used, including the 75\,m
Lovell telescope, providing baselines from 11--217 km (90--3910
k$\lambda$).  The data were processed in dual polarization, using
4$\times$128 MHz spectral windows, each divided into $64\times 2$\,MHz
channels.

The point-like QSO 0551+0829, separation $\sim1.\!^{\circ}5$, was used as
the phase reference on a cycle of 7:3 minutes. OQ208 was used as the
bandpass and flux density calibrator. The flux scale is
accurate to $\sim$10\%. The calibrated and edited
Betelgeuse data comprised 4--8\,hr per antenna, spread over 10.5\,hr,
with an average bandwidth of 400\,MHz. The data were imaged using two
schemes, one for highest resolution and the other, with a slight
tapering of the visibilities, for maximum sensitivity to the extended
stellar atmosphere, using a 180-mas circular restoring beam. The
latter was used to measure the astrometric position.  The
astrometric uncertainty arising from phase referencing is 16\,mas, and
adding the noise-based fitting uncertainty gives a total of 17.5\,mas.
\citet{2013MNRAS.432L..61R} describes these observations in more detail,
but note that the measurements given in this paper are after re-processing
using the correct receiver mount position for the Cambridge antenna \citep{Beswick2015}.

Similar observations were made in 2015 March and June, but the phase
reference 0605+0939 was used, $\sim7^{\circ}$ from the target, giving
a total astrometric uncertainty of 22.2 and 22.5\,mas in March and
June, respectively.

The total flux densities at successive epochs were 2.78, 2.39 and 2.35
mJy in areas 204$\times$195, 212$\times$198 and 201$\times$189 mas$^2$,
respectively, which is consistent with the barely significant ellipticity measured in VLA
data.  At all epochs, the high-resolution images (as reprocessed, for
2012) show different distributions of six to eight small hot spots 
at the level of $\approx 10\%$ of the total flux density. We tested whether these were
biasing the fits by subtracting the hot spots from the 2012 visibility
data and replacing them with the 2015 March hot spots, and re-imaging
the modified 2012 data at 180-mas resolution. The position fitted was
less than 1\,mas different from the original measurement, suggesting that
the hot spots seen at 5.75\,GHz do not significantly bias the 
(tapered) measured e-MERLIN centroid position and thus are not a significant source of {\it Cosmic Noise} for this analysis.

\subsection{Sources of Additional Radio Positional Noise}

In \citet{2008AJ....135.1430H} the radio positions were assigned positional uncertainties under the assumption that the intensity weighted center was the position of the barycenter. However, the \cite{Lim_1998Natur.392..575L} VLA Q-band (7\,mm, 43\,GHz) map shows the star is not axisymmetric being better fit with a $95 (\pm 2) \times 80 (\pm2)$\,mas ellipse with a position angle (PA) of $67 (\pm 7)\deg$ (East of North). The presence of non-axisymmetric intensity distribution directly implies that one must consider the possibility of an offset between the weighted intensity
centroid and the star's barycenter.  Elliptical fits have been made in subsequent multi-epoch multi-wavelength VLA maps \citep{OGorman_2015A&A...580A.101O} and the orientation
does not appear to reflect a fundamental property of the star, e.g., rotational distortion \citep{T2007ApJ...670L..21T}, and furthermore, the PA varies with time. 
For example, an elliptical fit to the ALMA high-spatial resolution map ($\sim 330$\,GHz), which samples closer to the photosphere than the VLA Q-band, gives 
$54.50 (\pm 0.01) \times 50.44 (\pm 0.01)$\,mas with a PA$=155.5 (\pm 0.2)\deg$
\citep{OGorman_2017}.

The problem then is how do we quantify the magnitude of the intensity centroid shift with respect to the barycenter? At this point in time nothing rigorous can be done
because reliable grids of physically complete 3D atmospheric models describing this 
part of the atmosphere do not exist. \citet{2011A&A...528A.120C} considered optical photo-center motions from radiative hydrodynamic simulations
of a red supergiant like Betelgeuse that uses only gray opacities in the visible surface layers. They find a dispersion of 0.5\,mas which is consistent with ground-based interferometry obtained by \citet{W1992MNRAS.257..369W} (0.4\,mas) and \citet{T1997MNRAS.285..529T} (1.2\,mas), but still smaller than the \hipp {\it Cosmic Noise} of 2.4\,mas. More recently, \citet{2016A&A...588A.130M} have presented H-band interferometry that reveals larger photo-center displacements at 4 epochs over 3 years with a mean displacement of 1.2\,mas. Furthermore, during 2014, the photo-center moved by 3.4\,mas. The H-band covers the 1.6$\mu$m H$^-$ opacity minimum and samples the deeper, more convective layers. SPHERE/ZIMPOL images from the VLT, with a resolution of $\sim 20$\,mas, also reveal non-symmetric photospheric emission in the V ($\lambda 554$\,nm; $\Delta\lambda=81$\,nm) and TiO717 filters ($\lambda 717$\,nm; $\Delta\lambda=20$\,nm) \citep{2016A&A...585A..28K}. The global asymmetries observed with SPHERE/ZIMPOL can introduce apparent photo-center offsets of order $\sim 1$\,mas.

For the high-spatial and high signal-to-noise ratio ALMA 2015 map there are residuals to the uniform intensity ellipse fits. Eight different fits to the ALMA data; an ellipse with additional Gaussian spots, rings, and point sources, lead to shifts in the apparent ellipse centroid with a standard deviation of 0.43\,mas. If we take these fits as possible realizations of radio-center displacements, then we can take 0.43\,mas as a measure of the minimum of additional radio noise - that might correspond to the \hipp {\em Cosmic Noise}.  However, the ratio of major to minor axis length of 1.1 suggests the potential for larger radio-center offsets. If we take the minor axis as representative
of an unperturbed star then there is a 5\,mas extension for the other axis. We adopt approximately half this value 
as the nominal radio noise, i.e. $\sigma_{radio}^{noise}=2.4$\,mas. The similarity to the \hipp {\em Cosmic Noise} in the 2007 solution may be coincidental because the sensitivity of specific intensity to changes in gas temperature are much smaller on the Rayleigh-Jeans tail of the Planck function.  Including this additional uncertainty in the radio positional errors (in quadrature) only has an effect on the few high-spatial resolution radio observations, which, apart from one, are from the VLA.

Simulations of optical photo-center displacement show time-scales of months and years
\citep{2011A&A...528A.120C}, so the photo-center displacements are not random, as assumed in the \hipp astrometric fitting procedures. However, at this point there is little more that we can do except to point out a potential future solution in 
\S\ref{lab:strategy}.

\section{Radio-Only Proper Motions and Parallax}

It has already been noted that VLA radio positions suggest different proper motions
to the 1997 \hipp astrometric solution \citep{B2003AJ....126..484B,2008AJ....135.1430H} [see Table~\ref{tab:propermotion_alt}]. The differences become more significant now because the 2007 \hipp solution has uncertainties that are a factor of $>2$ smaller.

Here, we consider the new radio dataset and explore the proper motions and parallaxes that result from different levels of additional radio centroid noise. 
We compute the parallaxes using the JPL DE405 IRCF ephemerides \citep{Standish1998}. 
These proper motions are presented in Table~\ref{tab:radio_only} where the uncertainties are calculated assuming the adopted astrometric model is the correct one. 
The parallaxes are not tightly constrained because there are only a few positions with
small positional uncertainties comparable to the size of the parallax. The parallaxes are smaller than given by either the 1997 and 2007 \hipp solutions and the parallax increases with increasing radio {\it Cosmic Noise}. Considering Table~\ref{tab:propermotion_alt}, it can be seen that
none of the proper motions ($\mu_{\alpha\cos\delta}$ or $\mu_\delta$) agree with the 2007 \hipp solution. This suggests that there is a bias in \hipp proper motions that may arise from too much weight being given to the direction $29^\circ$ E. of N. The PA of Betelgeuse's rotation axis, 
as deduced from combined UV spectral measurements, is $\sim 65^{\circ}$ E. of N. \citep{2006ApJ...646.1179H, 1998AJ....116.2501U} and with the a 
period of $\ge 17$\,yr equatorial motion of bright regions is unlikely to be an explanation. However, 1991-1992 interferometry
from \citet{W1992MNRAS.257..369W} and \citet{T1997MNRAS.285..529T} show hot spots which shift the photo-center with respect to the extended disk
to PA's of $39^\circ$ (1991 January) and $-29^\circ$ (1992 January) which can be attributed to photospheric convective motions \citep{2011A&A...528A.120C}.
This provides evidence that photo-center motions did occur during the \hipp mission, and these may have contributed to the necessity of adopting a stochastic solution.

\section{Revised Intermediate Astrometric Solutions}

The HIAD were retrieved from the DVD  included in \citet[Appendix G2.1]{2007ASSL..350.....V}\footnote{There is a very small difference in some of the adopted astrometric values available on-line and on the DVD.}. There are 66 scan level abscissa $(a)$ measurements, and we reject one outlier
measurement as in the 2007 published solution. This represents an increase in the number of measurements from the
original reduction (38). \citet[Eq.~2.52]{2007ASSL..350.....V} gives the differences  between predicted and measured abscissae (residuals, d$a$) as
\begin{equation}
{\rm d}a=\cos\psi\,{\rm d}\alpha\cos\delta + \sin\psi\, {\rm d}\delta + {\rm PARF}\,{\rm d}\pi +\cos\psi\,\Delta t\, {\rm d}\mu_{\alpha\cos\delta} + \sin\psi\,\Delta t\, {\rm d}\mu_\delta
\label{eq:da}
\end{equation}
where $\pi$ is the parallax and PARF (known as the parallax factor) $\psi$ is related to the scan angle. For each scan, the estimated abscissa uncertainties  $\sigma_a$ are also provided.
For the stochastic solution $\sigma_a$ includes the additional {\it Cosmic Noise} of $\sigma_{Hipp}^{noise}=2.4$\,mas. 
The residuals from the \citet{2007ASSL..350.....V} solution, d$a$, PARF, $\psi$, and $\Delta t$ are given in the DVD. The published solution represents the minimum of

\begin{equation}
\chi^2 = \sum \limits_{i=1}^{65}\left({{\rm d}a\over{\sigma_a}}\right)^2
\label{eq:chi2}
\end{equation}

Introducing information from the radio positions, e.g. proper motions, leads to perturbations about the 2007 \hipp solution, i.e., d$\alpha\cos\delta$, d$\delta$, d$\pi$, d$\mu_{\alpha\cos\delta}$, and d$\mu_\delta$ whose effect on the optimum solution can then be explored using Eqs~\ref{eq:da} and \ref{eq:chi2}. Both the radio and 2007 \hipp solutions are given in the International Celestial Reference System (ICRS). The uncertainty in the link between the 1997 Hipparcos catalog \citep{1997A&A...323L..49P} and ICRS at J1991.25 was 25\,\masyr in the relative motions of the frames, and the \hipp 2007 catalog is in agreement
to within the uncertainties \citep[S3.6]{2007ASSL..350.....V}.

\section{Results}

\subsection{Radio Fixed Proper Motions and HIAD}

The long time span of the radio observations provides a powerful constrain on the proper motions. By fixing $\mu_{\alpha\cos\delta}$ and $\mu_\delta$ for the case of $\sigma_{radio}^{noise}=2.4$\,mas (Table 2) in Eqs~\ref{eq:da} and \ref{eq:chi2}
we find a parallax of $\pi=5.27\pm 0.78$\,mas. This gives a distance of 
$190^{+33}_{-25}$\,pc which is larger than the \hipp revised 2007 value of $152^{+22}_{-17}$\,pc. Examination of the correlation coefficients of the 2007 \hipp solution shows that the parallax is positively correlated with both proper motions. The radio proper  motions are both smaller than the 2007 \hipp solution
so the parallax is also smaller, and the distance greater.

\subsection{Radio Positions and HIAD}

\begin{figure}[t]
\vspace{-2cm}
\includegraphics[angle=90,scale=0.75]{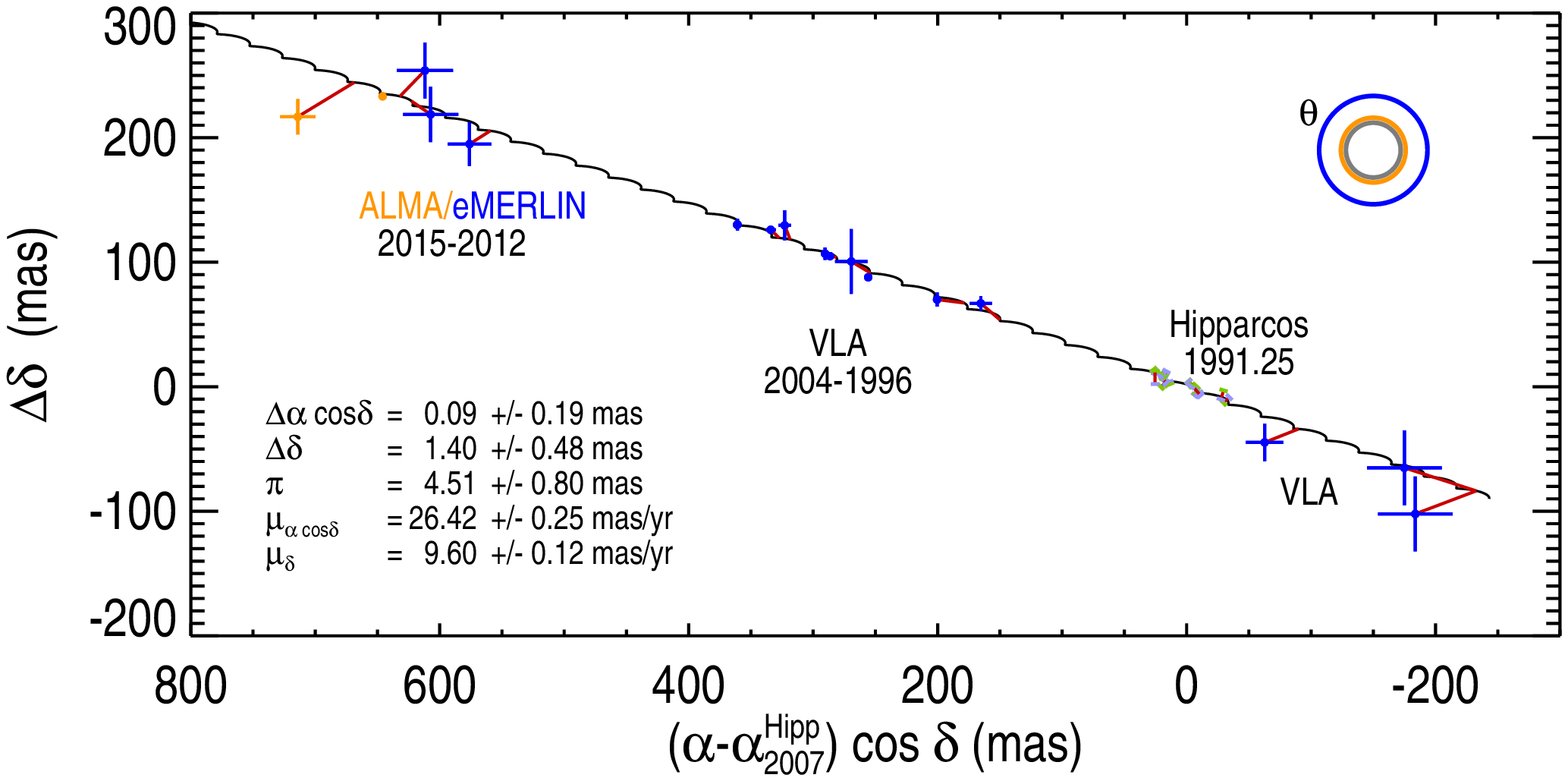}

\vspace{-5cm}
\includegraphics[angle=90,scale=0.75]{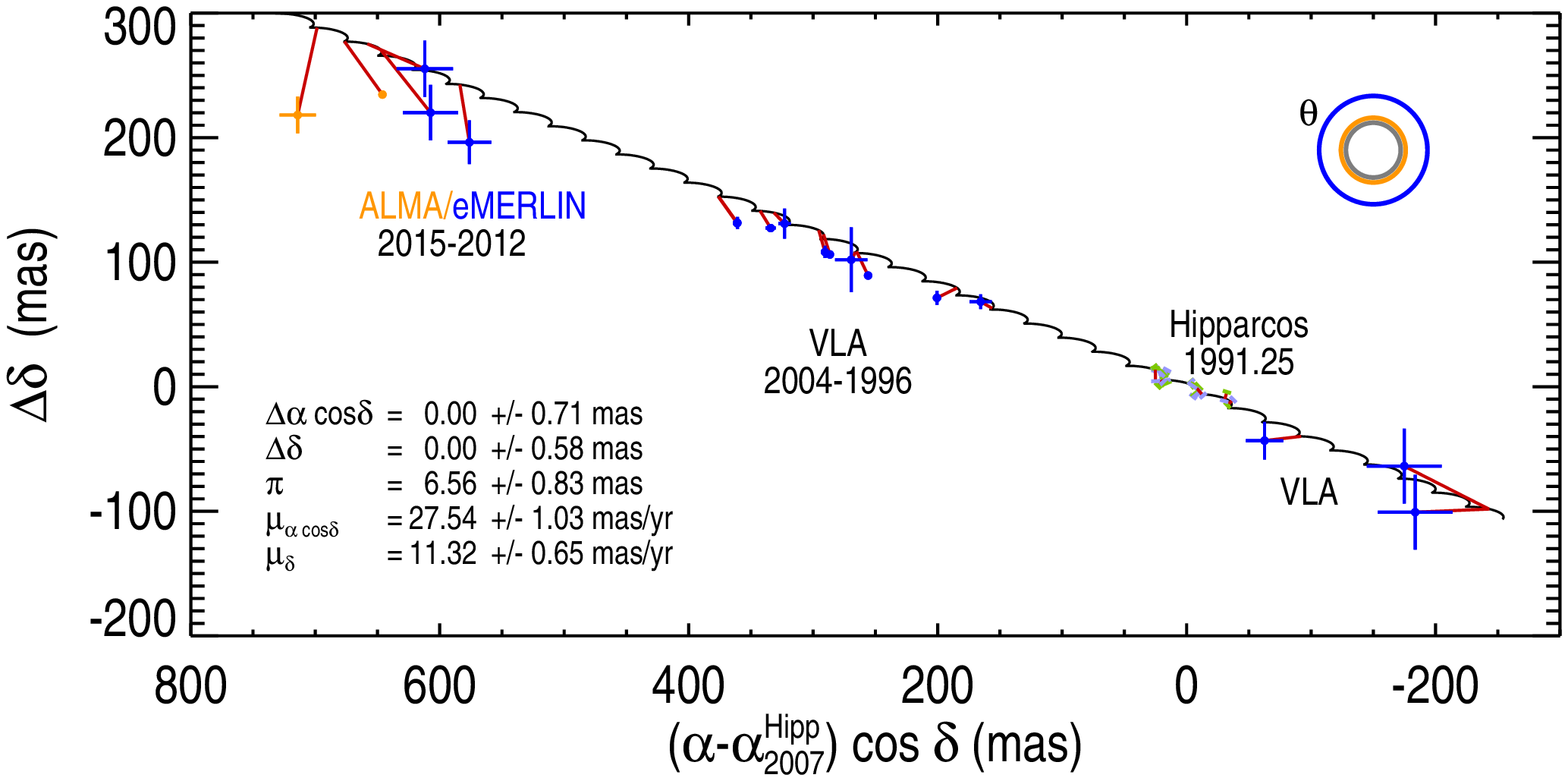}
\vspace{-2cm}
\caption{Top: The new astrometric solution computed with an assumed radio {\it Cosmic Noise} of 2.4\,mas. The 1997 HIAD Hipparcos data are shown for illustration \citep[see][]{2008AJ....135.1430H}. On this scale the 1997 \hipp data are not distinguishable from the 2007 data. The apparent angular diameter of the star at different wavelengths is shown in the upper-right corner for reference -- inner circle is H-band ($1.6\mu$m), the yellow is ALMA (338\,GHz) and blue is Q-band (43\,GHz). Bottom: The 2007 \hipp solution which shows
how the radio data constrains the proper-motions. The radio data show that \hipp systematically overestimates $\mu_\delta$ and to a lesser extent $\mu_{\alpha\cos\delta}$.}
\label{fig:soln}
\end{figure}

New astrometric solutions were then obtained using the revised HIAD combined with the full radio positional data in Table~\ref{tab:positions} and \citet[and references therein]{2008AJ....135.1430H}. We minimized the $\chi^2$ from the residuals in Eqs.~\ref{eq:da} and \ref{eq:chi2} combined with the radio position offsets ($d\alpha\cos\delta$ and $d\delta$) weighted by their uncertainties. We used Levenberg-Marquardt least-squares minimization \citep{2009ASPC..411..251M} with the 65 1D HIAD scans and each of the 18 2D radio position. For the radio, the residuals in 
$d\alpha\cos\delta$ and $d\delta$ were evaluated separately, effectively giving 36 datum ($2\times 18$ positions), for a total of 101 individual evaluations.  The result, assuming $\sigma_{radio}^{noise}=2.4$\,mas, is given in Table~\ref{tab:results}, along with the 2007 \hipp solution. For convenience, the position offsets for the 1991.25 Epoch are given with respect to the 2007 \hipp solution. Figure~\ref{fig:soln} (top) depicts this solution along with the radio positions, while the bottom panel shows the 2007 \hipp solution.

\begin{figure}
\epsscale{1.0}
\plotone{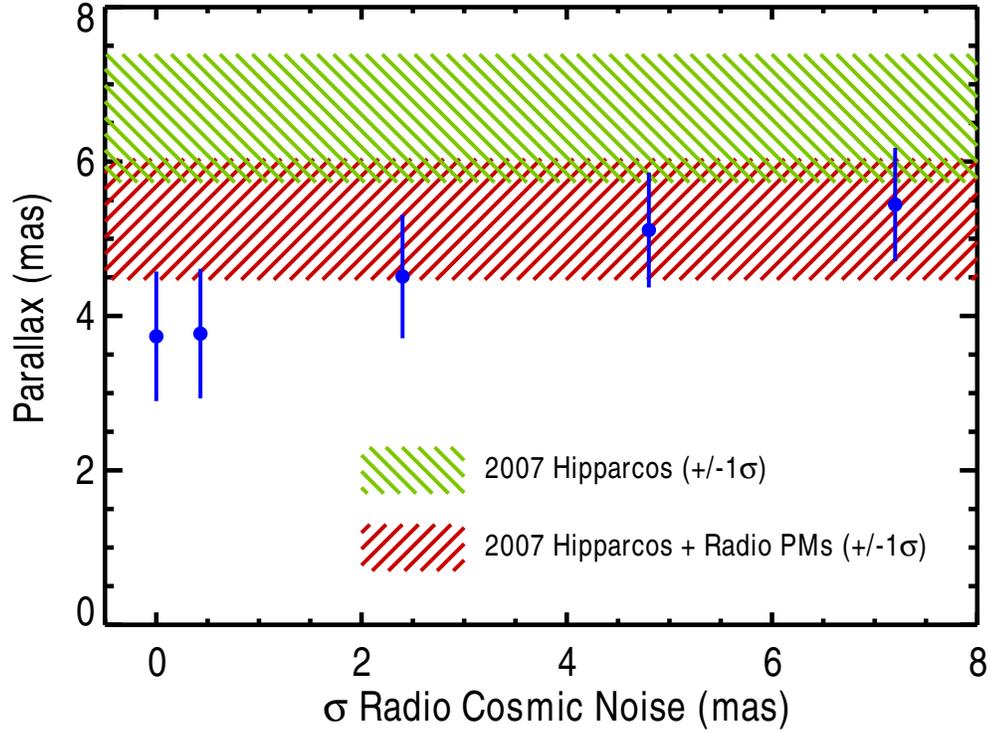}
\caption{The parallax from the combined radio and 2007 \hipp HIAD data as a function of {\it Radio Cosmic Noise} ($\sigma^{noise}_{radio}$). The top hashed area (green) is the 2007 Hipparcos solution, and the hashed area below (red) is \hipp HIAD combined with the radio-only proper motions (this work, assuming $\sigma^{noise}_{radio}=2.4$\,mas). The combined radio+\hipp
solutions are shown as circular symbols and error bars (blue) for different values of assumed $\sigma^{noise}_{radio}$. When $\sigma^{noise}_{radio}> 3$\,mas, the sensitivity of the parallax to the radio positions is reduced.}
\label{fig:radionoise}
\end{figure}

The most significant difference between the radio+\hipp and the 2007 \hipp solution is the proper motion in declination where the 2007 value of
$11.32\pm 0.65$ compares to $9.60\pm 0.12$\,mas/yr. This illustrates the importance of the long time span of the radio observations, and again 
hints that there is a preferred direction of the effective optical photo-center displacement during the \hipp observing epoch.   Astrophysically, the most important result is that the combined radio+\hipp
parallax is $4.51\pm 0.80$\,mas or $222^{+48}_{-33}$\,pc, which is $0.7\sigma$ greater than that found by Harper et al. (2008). Both the
solution using just the radio proper motions, and the combined radio+\hipp lead to a smaller parallax and greater distance than either of the
published \hipp results. Figure~\ref{fig:radionoise} shows the sensitivity of the new parallax to the adopted radio {\it cosmic noise}. Above 2.4\,mas the
few precise radio positions lose their impact and the combined solution is controlled by the proper-motion constraints (which remain insensitive to the adopted value of $\sigma_{radio}^{noise}$). 
There is certainly an argument to be made that the parallax could be smaller by adopting a smaller $\sigma_{radio}^{noise}$ in the combined solution.

The tension between the optical (closer distance) and radio (farther distance) means that any changes in the relative weighting of the two datasets, either through the number of effective measurements or the additional added {\it Cosmic Noise}, will shift the parallax to some degree. However, these results show that \hipp still overestimates the parallax. 

We note that Betelgeuse is very bright, too bright for {\it GAIA} to observe, and it is not clear that the small Japanese Nano-Jasmine 5.25\,cm IR telescope satellite\footnote{https://directory.eoportal.org/web/eoportal/satellite-missions/n/nano-jasmine} will be able to
get useful data for Betelgeuse (if it is launched).

\section{Discussion}

While the astrometric solutions presented above are probably the best that can be done at the present time, it is not the final solution to the problem
of the stellar distance. The present parallax for Betelgeuse implies it is 
1.2 times more luminous than in \citet{2008AJ....135.1430H}, but the uncertainties remain significant for astrophysical purposes.
To obtain a solid parallax measurement requires multiple observations where one could simultaneously measure the intensity position, with reference to accurate position (phase) calibrators, {\em and} image the uneven intensity distribution on the stellar surface. In this way, one has the chance that the continuum centroid and uncertainties from surface intensity variations could be estimated with sufficient precision to enable
a reliable parallax determination. 

\subsection{What is the optimum wavelength to make astrometric measurements?}
\label{lab:strategy}

The outer atmosphere of the convective M supergiant may be shaped by gas pressure,
rotation, convective motions, shocks, and magnetic fields competing with the stellar gravitational field. These may lead to surface brightness variations that are not centered on the star's barycenter which in turn will influence the astrometric solution. 
For Betelgeuse, rotation is not expected to lead to significant surface distortion \citep{T2007ApJ...670L..21T}. One would like to sample the layers well above the optically-thin visual surface and where convection-induced height variations are minimized. Surface magnetic fields are expected to decrease more slowly with height than the gas density (pressure) and magnetic fields are expected to shape the outer layers \citep{2013EAS....60...59H}, including the hotter ultraviolet emitting chromospheric region near 2$R_\ast$. When the gas temperature increase to above $\sim 4000$\,K, hydrogen becomes a significant electron donor and the radio opacity can increase significantly. This suggests that an optimal radial location for astrometric measurements is well above the photosphere but interior to the higher temperature chromosphere. This region is still relatively uncharted and is expected to host low-temperature inhomogeneities related to the MOLsphere phenomenon \citep{2006ApJ...645.1448T, O2009A&A...503..183O}, which is probably the cooler and lower-gravity equivalent of the COmosphere phenomenon observed in warmer stars, including K giants and the Sun \citep{1994ApJ...423..806W, 2002ApJ...575.1104A}. 

The intensity changes ($\Delta I$) resulting from differences in temperature ($\Delta T$) in a given layer are expected to be less in the sub-mm and cm than in the optical because the thermal continuum emission is on the Rayleigh-Jeans tail where perturbations should be linear in temperature, i.e., $\Delta I\propto \Delta T$. In the optical, the sensitivity is greater because of the
exponential nature of the Planck function, and in the V-band there is additional sensitivity resulting from the sensitivity of molecular opacity to temperature. 

Fortunately, the mm and sub-mm radio frequencies probe the optimum atmospheric layers, and in addition, high-spatial resolution observations can image these layers at the same time as the position is accurately known with respect to an external reference frame (ICRS).  This may be achieved with ALMA in the sub-mm, and perhaps in combination with the JVLA at Q-band (43\,GHz, 6.9\,mm). These observations would need maximum baseline configurations for the highest spatial resolution. 

To obtain an accurate and precise parallax for Betelgeuse will require a dedicated multi-year observing plan. In addition to requiring good weather conditions for atmospheric phase stability, Betelgeuse observations with multiple check sources are needed near maximum parallax shift. Redundant observations will need to be planned because experience shows that data will be lost to bad weather. Another challenging constraint is that interferometer array configurations are routinely changed in non-yearly cycles to satisfy other science priorities.

\section{Conclusions}

The VLA+ 1997 \hipp HIAD astrometric solution for Betelgeuse \citep{2008AJ....135.1430H} has been improved by using the revised 2007 \hipp HIAD from \citet{2007ASSL..350.....V} combined with extant VLA data and 
new e-MERLIN and ALMA positions. The non-axisymmetric intensities found in 
high-spatial resolution VLA and ALMA radio maps indicate that the intensity centroid may not coincide with the barycenter, leading us to introduce radio {\it Cosmic Noise} analogous to that required for the \hipp optical solutions.

First, we find that using the proper motions from the radio data alone, which has a time span of 34~years, leads to a smaller parallax (greater distance)
than either \hipp solution (time span of 28 months). Second, the new combined radio+\hipp astrometric solution gives a parallax and distance of $4.51 \pm 0.80$\,mas 
and $222^{+48}_{-34}$\,pc, respectively. This distance is slightly larger, but not significantly so, than the previous radio+\hipp solution of $197^{+55}_{-35}$\,pc \citep{2008AJ....135.1430H}. However, the new solutions are significantly different from the 2007 \hipp solution of \citet{2007ASSL..350.....V}. To obtain a more precise
parallax will require high-spatial resolution continuum observations in the mm and sub-mm that can be obtained with ALMA and JVLA. Scheduling and obtaining the required observations may take a Herculean effort.

\acknowledgements
Financial support for this work was provided by NASA through a SOFIA grant NAS2-97001 issued by USRA to GMH, who also acknowledges the assistance of R.W.H.  P.K. acknowledges financial support from the Programme National de Physique Stellaire (PNPS) of CNRS/INSU, France. E.O.G. acknowledges support from the Irish Research Council.  L.D. acknowledges support from the ERC consolidator grant 646758 AEROSOL and the FWO Research Project grant G024112N.  This paper makes use of the following ALMA data: ADS/JAO.ALMA\#2015.1.00206.S. 
ALMA is a partnership of ESO (representing its member states), NSF (USA) and NINS (Japan), together with NRC (Canada), NSC and ASIAA (Taiwan), and KASI (Republic of Korea), in cooperation with the Republic of Chile. The Joint ALMA Observatory is operated by ESO, AUI/NRAO and NAOJ.  The National Radio Astronomy Observatory is a facility of the National Science Foundation operated under cooperative agreement by Associated Universities, Inc.  
This research has made use of the VizieR catalogue access tool, CDS, Strasbourg, France. It has also made use of NASA's Astrophysics Data System Bibliographic Services. Finally, we thank the referee for their insights.



{\it Facilities:} \facility{Hipparcos}, \facility{VLA}, \facility{ALMA}, \facility{e-MERLIN}.




\clearpage


\begin{table}
\begin{center}
\caption{New ICRS Radio Positions for Betelgeuse.}
\label{tab:positions}
\begin{tabular}{lcccc}
\tableline
Date        &  RA                 & Decl.               & Unc.   & Reference   \\
            & (h m s)             & ($^\circ$ $\>\>$ \arcmin $\>\>\>$ \arcsec)    & (mas)  &        \\ \tableline
2012.530    & 05 55 10.32672      &  07 24 25.5263     & 17.5   & 5.75\,GHz, \citet{2013MNRAS.432L..61R} \\
2015.178    & 05 55 10.32882      &  07 24 25.5501     & 22.2   & 5.75\,GHz, New \\
2015.422    & 05 55 10.32913      &  07 24 25.5853     & 22.5   & 5.75\,GHz, New \\
2015.854    & 05 55 10.33142      &  07 24 25.5646     &  2.0   & 330-345\,GHz ALMA\\ 
2016.623    & 05 55 10.33560      &  07 24 25.5482     & 14.1   & 330-345\,GHz ALMA\\ 
\tableline
\end{tabular}
\end{center}
\end{table}

\begin{table}
\begin{center}
\caption{Proper Motions for Betelgeuse\tablenotemark{a}. }
\label{tab:propermotion_alt}
\begin{tabular}{lcccc}
\tableline
Reference    & $\mu_{\alpha\cos\delta}$ & $\sigma_{\mu_{\alpha\cos\delta}}$ & $\mu_\delta$ & $\sigma_{\mu_\delta}$ \\\tableline 
ESA (1997)           & 27.33 & 2.30  & 10.86  & 1.46  \\
Boboltz et al. (2007)  & 23.98 & 1.04  & 10.07  & 1.15  \\
Harper et al. (2008)   & 24.95 & 0.08  &  9.56  & 0.15  \\
van Leeuwen (2007)     & 27.54 & 1.03  & 11.30  & 0.65  \\
Radio-only [$\sigma^{noise}_{radio}=2.4$\,mas] & 25.53 & 0.31  &  9.37 & 0.28  \\
\tableline
\end{tabular}
\tablenotetext{a}{Units: \masyr}
\end{center}
\end{table}

\begin{table}
\begin{center}
\caption{Radio-only Astrometric Solutions\tablenotemark{a}}
\label{tab:radio_only}
\begin{tabular}{lcccccc}
\tableline
$\sigma^{noise}_{radio}$ & $\pi$ & $\sigma_\pi$ & $\mu_{\alpha\cos\delta}$ & $\sigma_{\mu_{\alpha\cos\delta}}$ & $\mu_\delta$ & $\sigma_{\mu_\delta}$ \\\tableline 
0.00\,mas   & 3.33 & 1.93 & 25.77  & 0.28  & 9.55 & 0.24  \\
0.43\,mas   & 3.38 & 1.94 & 25.75  & 0.28  & 9.54 & 0.24  \\
2.40\,mas   & 3.77 & 2.20 & 25.53  & 0.31  & 9.37 & 0.28  \\ 
4.80\,mas   & 4.35 & 2.67 & 25.42  & 0.35  & 9.28 & 0.32  \\ 
7.20\,mas   & 5.16 & 3.11 & 25.37  & 0.37  & 9.26 & 0.35  \\
\tableline
\end{tabular}
\tablenotetext{a}{Units: mas for $\pi$, \masyr for $\mu$}
\end{center}
\end{table}

\begin{table}
\begin{center}
\caption{Astrometric Solutions for Betelgeuse.}
\label{tab:results}
\begin{tabular}{lcc}
\tableline
Parameter    &  van Leeuwen (2007) & Present Work     \\
\tableline
$\Delta\alpha\cos\delta$ (deg)  & 0.00(0.71)        & 0.09(0.19)     \\
$\Delta\delta$ (deg)  & 0.00(0.58)                  & 1.40(0.48)     \\
Parallax $\pi$ (mas)  & 6.55(0.83)                  & 4.51(0.80)     \\
$\mu_{\alpha\cos\delta}$ (\masyr) & 27.54(1.03)     & 26.42(0.25)    \\
$\mu_\delta$ (\masyr)  & 11.30(0.65)                & 9.60(0.12)    \\
\tableline
\end{tabular}
\end{center}
\end{table}

\bibliography{parallax_bibtex_2017_June13} 





\end{document}